\documentclass{article} 
\usepackage{longcat_style,times}

\usepackage{amsmath,amsfonts,bm}

\def\eqref#1{equation~\ref{#1}}

\def\1{\bm{1}}

\DeclareMathAlphabet{\mathsfit}{\encodingdefault}{\sfdefault}{m}{sl}
\SetMathAlphabet{\mathsfit}{bold}{\encodingdefault}{\sfdefault}{bx}{n}

\usepackage{hyperref}
\usepackage{url}

\usepackage{graphicx}
\usepackage{balance}  
\usepackage{amsmath}
\usepackage{algpseudocode}
\usepackage{latexsym}
\usepackage{multirow}
\usepackage{multicol}
\usepackage{color}
\usepackage{dashrule}
\usepackage{extarrows}
\usepackage{dsfont}
\usepackage{centernot}
\usepackage{hyperref}
\usepackage{natbib}
\usepackage{fancybox}
\usepackage[utf8]{inputenc} 
\usepackage[T1]{fontenc}    
\usepackage{hyperref}       
\usepackage{url}            
\usepackage{booktabs}       
\usepackage{amsfonts}       
\usepackage{amsmath}
\usepackage{nicefrac}       
\usepackage{microtype}      
\usepackage{xcolor} 
\usepackage{amsmath}
\usepackage{lipsum}
\usepackage{colortbl}
\usepackage{subcaption}
\usepackage{tabularx}
\usepackage{makecell}
\usepackage{wrapfig}
\usepackage{bm}
\usepackage{multirow}
\usepackage{xcolor} 
\usepackage{bbm}
\usepackage{mdframed}
\usepackage{enumitem}
\usepackage{xspace}
\usepackage{algpseudocode}
\usepackage{algpseudocode}
\usepackage[ruled,vlined,boxed]{algorithm2e}
\usepackage{algorithmicx}

\usepackage[most]{tcolorbox}
\usepackage{xcolor}
\usepackage{fancyvrb}
\usepackage{fvextra}

\usepackage[utf8]{inputenc} 
\usepackage{textcomp}       

\usepackage{framed}  

\DeclareUnicodeCharacter{00A0}{ }          
\DeclareUnicodeCharacter{2014}{-}          
\DeclareUnicodeCharacter{2013}{-}          
\DeclareUnicodeCharacter{200B}{}           

\definecolor{lightgray}{gray}{0.95}

\usepackage{lineno}

\definecolor{darkblue}{rgb}{0, 0, 0.5}
\hypersetup{colorlinks=true, citecolor=darkblue, linkcolor=darkblue, urlcolor=darkblue}

\usepackage{listings}
\usepackage{xcolor}
\usepackage[ruled,vlined]{algorithm2e}
\usepackage{caption}
\usepackage{float} 
\usepackage{arydshln}
\usepackage{wrapfig} 
\usepackage{tcolorbox}
\tcbuselibrary{theorems}
\tcbuselibrary{most}
\usepackage[dvipsnames]{xcolor}

\usepackage{hyperref}       
\usepackage{url}            
\usepackage{booktabs}       
\usepackage{amsfonts}       
\usepackage{nicefrac}       
\usepackage{microtype}      
\usepackage{xcolor}         
\usepackage{amsmath}
\usepackage{multirow}
\usepackage{multicol}
\usepackage{colortbl}
\usepackage{tcolorbox}
\usepackage{wrapfig}

\usepackage{cleveref}
\usepackage{xspace}

\makeatletter
\DeclareRobustCommand\onedot{\futurelet\@let@token\@onedot}
\def\@onedot{\ifx\@let@token.\else.\null\fi\xspace}

\makeatother

\usepackage{natbib}

\definecolor{light-gray}{gray}{0.6}
\definecolor{front-color}{HTML}{F5FFFA}
\definecolor{Gray}{gray}{0.93}

\definecolor{customTeal}{RGB}{0, 128, 128} 
\definecolor{emphasisColor}{RGB}{255, 0, 0} 
\definecolor{oursBlue}{RGB}{51,202,246}

\definecolor{blue1}{HTML}{508AB2}
\definecolor{green2}{HTML}{BFF6BA}

\newtcbtheorem[]{prompt}{Prompt}%
{colback=SeaGreen!10!CornflowerBlue!10,
 colframe=RoyalPurple!55!Aquamarine!100!,
 fonttitle=\bfseries,
 left=.02in, right=.02in, bottom=.02in, top=.02in,
 before upper={\linespread{1.5}\selectfont}}{prompt}

\renewcommand{\shorttitle}{OpenGame: Open Agentic Coding for Games}

\definecolor{darkblue}{rgb}{0, 0, 0.5}
\hypersetup{colorlinks=true, citecolor=darkblue, linkcolor=darkblue, urlcolor=darkblue}

\makeatletter
\renewcommand{\@maketitle}{%
  \vbox{%
    \hsize\textwidth
    \linewidth\hsize
    \vskip -0.5in
    \noindent
    \begin{minipage}{0.99\textwidth}

  \includegraphics[width=0.28\linewidth]{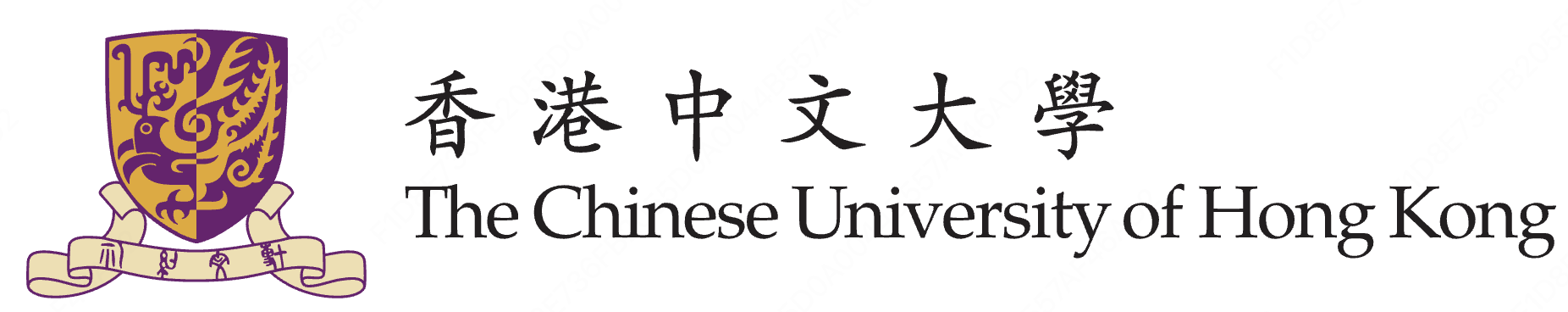} 
    \end{minipage}%
    \\
    \rule{\linewidth}{1pt}
    \hspace{0.05\textwidth}%
    \begin{minipage}{0.8\textwidth}
    \end{minipage}

\vskip -0.1in
    \centering
    {\LARGE \bfseries\@title\par}
    \vskip 0.15in  
    \def\And{%
      \end{tabular}\hfil\linebreak[0]\hfil%
      \begin{tabular}[t]{c}\bf\rule{\z@}{24\p@}\ignorespaces%
    }
    \def\AND{%
      \end{tabular}\hfil\linebreak[4]\hfil%
      \begin{tabular}[t]{c}\bf\rule{\z@}{24\p@}\ignorespaces%
    }
    \begin{tabular}[t]{c}\bf\rule{\z@}{24\p@}\@author\end{tabular}%
  \vskip 0.05in 
  }
}
\makeatother

\title{OpenGame: Open Agentic Coding for Games\\}

\makeatletter
\def\@fnsymbol#1{\ensuremath{\ifcase#1\or \dagger\or \ddagger\or
   \mathsection\or \mathparagraph\or \|\or **\or \dagger\dagger
   \or \ddagger\ddagger \else\@ctrerr\fi}}
\makeatother

\newcommand{\homepage}{\raisebox{-1.5pt}{\includegraphics[height=1em]{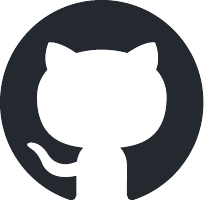}}}

\author{%
  \textbf{Yilei Jiang \quad Jinyuan Hu \quad Qianyin Xiao \quad Yaozhi Zheng \quad Ruize Ma \quad Kaituo Feng} \\
  \textbf{Jiaming Han \quad Tianshuo Peng \quad Kaixuan Fan \quad Manyuan Zhang \quad Xiangyu Yue\thanks{Corresponding author.}} \\
  \\
  CUHK MMLab \\
  \texttt{yljiang@link.cuhk.edu.hk, xyyue@ie.cuhk.edu.hk} \\[1ex]
  {\homepage\ \normalfont \texttt{Project Page: \url{https://yelonhu.github.io/OpenGame-landing-page/}}} \\
  {\normalfont \texttt{GitHub: \url{https://github.com/leigest519/OpenGame}}}
}

\begin{document}

{%
   \renewcommand\twocolumn[1][]{#1}%
   \maketitle
   \vspace{-1pt}
   \begin{center}
    \centering
    \includegraphics[width=0.99\linewidth]{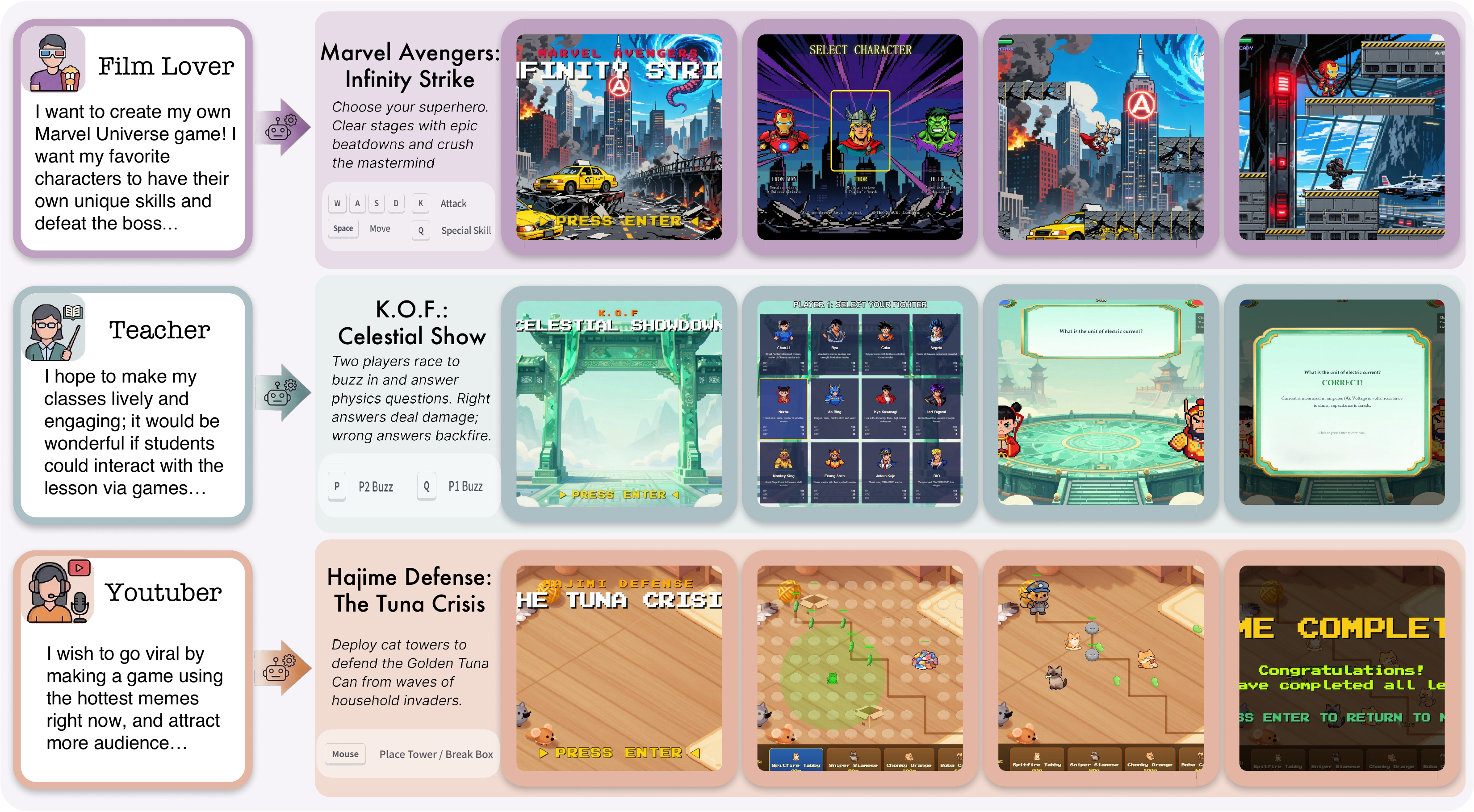
    }
    \captionof{figure}{\textbf{End-to-end agentic game generation with OpenGame.} Diverse users provide natural language specifications to autonomously create fully playable 2D games across distinct genres (e.g., action, educational, and tower defense). Each generated project features a complete game lifecycle seamlessly integrated with multimodal visual and audio assets.}
    \vspace{6pt}
   \end{center}%
  }
\begin{abstract}

Game development sits at the intersection of creative design and intricate software engineering, demanding the joint orchestration of game engines, real-time loops, and tightly coupled state across many files. While Large Language Models (LLMs) and code agents now solve isolated programming tasks with ease, they consistently stumble when asked to produce a fully playable game from a high-level design, collapsing under cross-file inconsistencies, broken scene wiring, and logical incoherence. We bridge this gap with \textbf{OpenGame}, the first open-source agentic framework explicitly designed for end-to-end web game creation. At its core lies \textbf{Game Skill}, a reusable, evolving capability composed of a \textit{Template Skill} that grows a library of project skeletons from experience and a \textit{Debug Skill} that maintains a living protocol of verified fixes—together enabling the agent to scaffold stable architectures and systematically repair integration errors rather than patch isolated syntax bugs. Powering this framework is {GameCoder-27B}, a code LLM specialized for game engine mastery through a three-stage pipeline of continual pre-training, supervised fine-tuning, and execution-grounded reinforcement learning. Since verifying interactive playability is fundamentally harder than checking static code, we further introduce {OpenGame-Bench}, an evaluation pipeline that scores agentic game generation along Build Health, Visual Usability, and Intent Alignment via headless browser execution and VLM judging. Across 150 diverse game prompts, OpenGame establishes a new state-of-the-art. We hope OpenGame pushes code agents beyond discrete software engineering problems and toward building complex, interactive real-world applications. Our framework will be fully open-sourced.

\end{abstract}    
\section{Introduction}
\label{sec:intro}

Video games represent one of the sharpest challenges in automated software engineering, demanding a rare fusion of rigorous logic, aesthetic design, and interactive storytelling. Unlike traditional utility software, a playable game is a real-time system whose quality depends on the seamless orchestration of update loops, physics, event handling, asset pipelines, and tightly coupled state across many files. This makes game creation both technically demanding and creatively expensive. Although democratizing game development has long been a goal of the broader creative community, the barrier to entry remains stubbornly high: turning an idea into a playable artifact still requires simultaneous mastery of engine architectures, programming languages, and fragile systems integration.

The recent surge in Large Language Models (LLMs) and autonomous code agents has transformed the landscape of software engineering~\cite{jimenez2023swe, devin}. Modern agents can solve discrete algorithmic problems, generate boilerplate, and even navigate mature repositories with impressive competence. Yet when tasked with end-to-end game creation, these general-purpose systems hit a formidable ``complexity wall.'' Generating a calculator script or an isolated gameplay mechanic is far easier than constructing a coherent, fully playable game. In practice, we observe three recurring failure modes in frontier models: (1) \textbf{Logical Incoherence}: the model loses track of global state across the game loop, producing projects that freeze, fail to terminate, or never realize key mechanics; (2) \textbf{Engine-Specific Knowledge Gaps}: general models often ignore or misuse engine abstractions, re-implementing mechanics from scratch instead of correctly leveraging framework-native physics, scene, and event systems; and (3) \textbf{Cross-File Inconsistencies}: even when individual files look plausible, the overall project frequently breaks due to mismatched asset keys, flawed scene wiring, missing configuration fields, or broken initialization order. These failure modes are precisely what prevent natural-language game design from being reliably brought to life.

To bridge this gap, we argue that the field must move beyond generalist code agents toward specialist frameworks that understand the intrinsic structure of games. We therefore present \textbf{OpenGame}, the first open-source agentic framework explicitly designed for end-to-end web game creation. At the core of OpenGame is \textbf{Game Skill}, a reusable capability for translating a natural-language design specification into a runnable project. Game Skill addresses systemic integration failures through two evolving components. First, \textbf{Template Skill} grows an evolving library of project skeletons ($\mathcal{L}$), starting from a single game-agnostic meta template ($\mathcal{M}_0$) and expanding into specialized template families such as gravity-based side view and top-down continuous motion. This sharply reduces the search space of generation and stabilizes project-wide structure. Second, \textbf{Debug Skill} maintains a living debugging protocol ($\mathcal{P}$) updated from observed build, test, and runtime outcomes, allowing the agent to accumulate verified fixes and systematically resolve high-frequency integration failures rather than repeatedly rediscovering them from scratch.

Supporting this framework is a domain-specialized foundation model, \textbf{GameCoder-27B}. Rather than relying solely on prompting a general code model, we train GameCoder-27B through a three-stage pipeline of continual pre-training, supervised fine-tuning, and execution-grounded reinforcement learning. This pipeline equips the model with engine-specific architectural priors, API usage patterns, and the logical discipline required for multi-file gameplay systems, providing a stronger substrate for the downstream agent.

Finally, progress in this area is bottlenecked by evaluation. Validating a game is fundamentally harder than verifying a standard function: code that compiles may still produce an unplayable, inert, or mechanically incoherent experience. Existing software benchmarks predominantly rely on static input-output unit tests, which are poorly suited to the temporal and interactive nature of gameplay. To address this gap, we introduce \textbf{OpenGame-Bench}, an evaluation pipeline designed to assess whether an agent can actually build interactive web games. OpenGame-Bench moves verification from static code analysis to dynamic playability assessment, scoring generated projects along build correctness, visual usability, and intent satisfaction through headless browser execution and multimodal judging.

In summary, our contributions are as follows:
\begin{itemize}
    \item We propose OpenGame, the first open-source, tool-augmented coding agent dedicated to generating playable web games from natural-language specifications, enabling creative design ideas to be brought to life as executable artifacts. Central to this framework is Game Skill, an evolving combination of Template Skill and Debug Skill that stabilizes project scaffolding and resolves recurrent cross-file failures.
    \item We train GameCoder-27B, a domain-specialized code model through continual pre-training, supervised fine-tuning, and execution-grounded reinforcement learning to better master game engine patterns, API usage, and complex gameplay logic.
    \item  We introduce OpenGame-Bench, a new evaluation paradigm for interactive code generation, moving beyond static unit tests to measure build health, visual usability, and intent alignment for end-to-end web game creation.
\end{itemize}
\section{Related Work}
\label{sec:related}

\textbf{Agentic Benchmarks and Software Development.}
Software development has become one of the premier frontiers for evaluating autonomous agents. SWE-Bench~\citep{jimenez2024swebenchlanguagemodelsresolve, yang2024sweagent} catalyzed the shift toward agentic software engineering by moving evaluation from isolated functions to repository-level issues. Since then, multiple software benchmarks have emerged to test complex unimodal reasoning~\citep{chan2025mlebenchevaluatingmachinelearning, merrill2026terminalbenchbenchmarkingagentshard, yang2025codeclashbenchmarkinggoalorientedsoftware}, while efforts to introduce multimodal capabilities have predominantly focused on frontend JavaScript development~\citep{Zhu2025FrontendBenchABA, Si2024Design2CodeHFA, yang2024swebenchmultimodalaisystems}. Beyond pure coding, multimodal agents are frequently evaluated on computer use~\citep{xie2024osworldbenchmarkingmultimodalagents} and web navigation~\citep{zhou2024webarenarealisticwebenvironment, koh2024visualwebarenaevaluatingmultimodalagents}. More recently, GameDevBench frames game development itself as a testbed for evaluating agentic capabilities~\citep{chi2026gamedevbenchevaluatingagenticcapabilities}. Progress in this space is especially challenging because agents must not only operate within multimodal action spaces, but also produce executable software artifacts whose quality unfolds over time during interaction. Game development therefore bridges two difficult regimes: it demands the multimodal grounding of computer-use agents, yet still requires the definitive code synthesis and structural consistency of software agents. {OpenGame-Bench} is designed for this exact intersection, focusing on end-to-end interactive web game construction and dynamic playability evaluation rather than static task completion alone.

\textbf{AI in Games: From Playing to Content Generation.}
Historically, games have served as interactive simulation environments and proxies for evaluating AI intelligence~\citep{gallotta2024large}, spanning seminal milestones like Deep Blue~\citep{campbell2002deep}, AlphaGo~\citep{silver2016mastering}, and Cicero~\citep{meta2022human}, to modern 3D generalists like SIMA 2~\citep{bolton2025sima}. Recently, a flux of agents designed to play games, such as LLMs navigating Pokémon~\citep{karten2025pokeagent, karten2025pok, comanici2025gemini, nunuai2024pokemon}, has explored the reasoning capabilities of frontier models and assisted in game testing. However, transitioning from non-player characters or testers to driving the actual game \textit{development} process introduces structural challenges. In the realm of game creation, AI has progressed from automated asset creation via Procedural Content Generation~\citep{summerville2018proceduralcontentgenerationmachine, shaker2016procedural} and evolutionary level design~\citep{sudhakaran2023mariogptopenendedtext2levelgeneration} to bypassing traditional engines entirely. Frameworks like Concordia substitute mechanics with LLM-driven adaptive stories~\citep{vezhnevets2023generative, vezhnevets2025multi}, while neural world models like Genie attempt to simulate physics and generate frames interactively~\citep{bruce2024genie}. 

\textbf{Structural Game Engineering and Web-Based Frameworks.}
While neural simulations push the boundaries of content generation, they are not aligned with how professional games are built, which requires deterministic game engines. However, industry-standard engines like Unreal Engine~\citep{unrealengine} and Unity~\citep{unity} rely heavily on proprietary GUIs and binary asset serialization, making them notoriously difficult for text-based autonomous agents. In contrast, web-based 2D frameworks like Phaser~\citep{phaser} provide a purely programmatic API surface highly amenable to LLMs. Because a complete Phaser game can be expressed entirely in raw JavaScript or TypeScript, it serves as an ideal testbed for agentic software engineering. It is here that \textbf{OpenGame} distinguishes itself from generating isolated assets or simulated frames. By targeting the text-driven architecture of the Phaser engine, our GameCoder-27B model employs a \textbf{Game Skill}—comprising an evolving \textit{Template Skill} for project scaffolding and a living \textit{Debug Skill} to resolve cross-file inconsistencies—to output verifiable, executable web games. Consequently, \texttt{OpenGame-Bench} evaluates exactly what is required to combine these features, setting a new standard for interactive code generation.
\section{Methodology}
\label{sec:methodology}

\begin{figure}[t]
    \centering
    \includegraphics[width=0.98\textwidth]{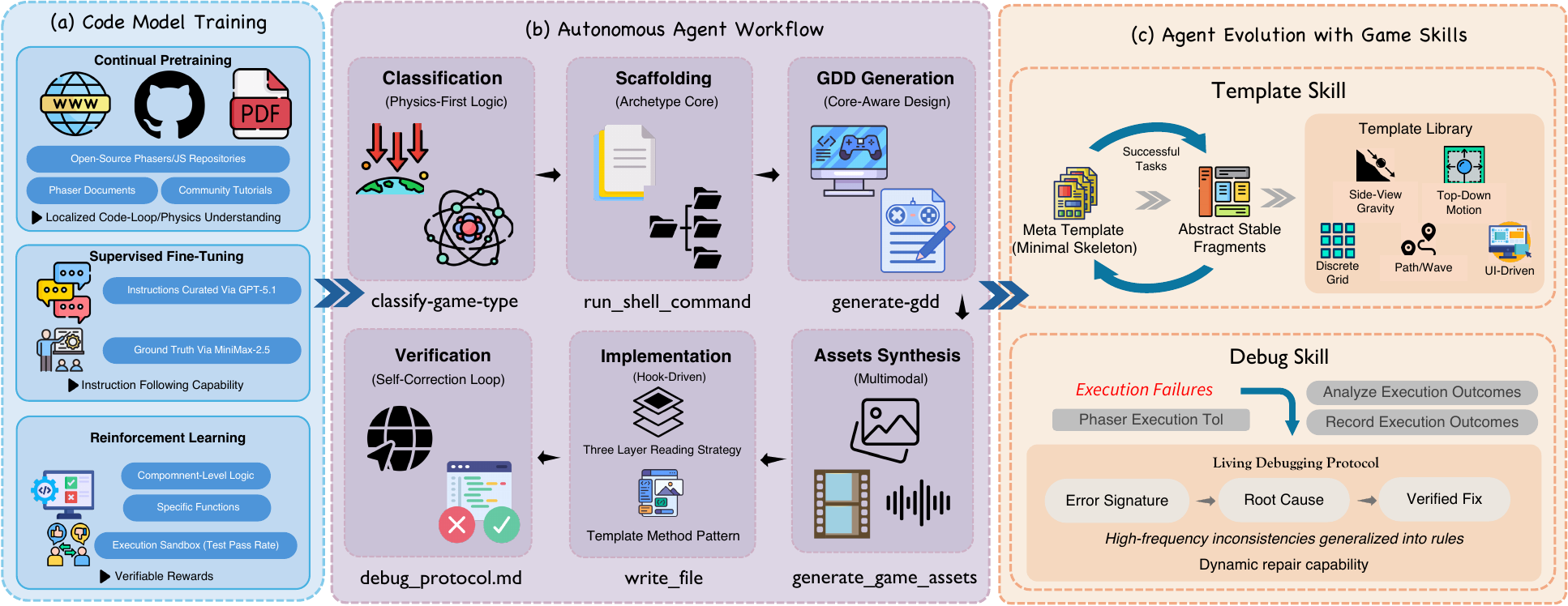}
    \caption{\textbf{The OpenGame architecture.} The framework integrates three coupled components: (a) a multi-stage code-model training pipeline that establishes engine-specific priors, (b) an autonomous agent workflow that translates natural-language game ideas into runnable projects through a structured six-phase process, and (c) an agent-evolution module that continuously refines structural scaffolding (Template Skill) and repair behavior (Debug Skill) through accumulated experience.}
    \label{fig:opengame_architecture}
\end{figure}

OpenGame is built from the interaction of a domain-specialized code model and a structured multimodal coding agent. Our methodology has three pillars: the training pipeline of the base model (GameCoder-27B), the design of the autonomous game-generation workflow, and the continual evolution of the agent through reusable game-development skills.

\subsection{Base Model Training}
To provide the foundational logic and engine-specific knowledge required for game development, we develop \textbf{GameCoder-27B}, built on top of a Qwen3.5-27B backbone. Standard LLMs often struggle to synthesize the multi-file structures required by engines such as Phaser. We address this gap through a three-stage training pipeline: Continual Pre-Training (CPT), Supervised Fine-Tuning (SFT), and Reinforcement Learning (RL).

\textbf{Continual Pre-Training (CPT):} We first adapt the base model to the domain of interactive web games. We assemble a large-scale pre-training corpus from open-source Phaser and JavaScript/TypeScript game repositories on GitHub, together with official documentation and community tutorials. This stage builds a strong prior over game loops, physics systems, asset usage, and state management patterns.

\textbf{Supervised Fine-Tuning (SFT):} To align the model with instruction-following for game design, we synthesize a diverse question-answer dataset. We leverage \textit{gpt-codex5.1} to curate complex, multi-step game design prompts (e.g., ``Implement a 2D platformer character controller with double-jump and sprite animations''). We then use \textit{minimax2.5} to produce high-quality target solutions. This synthetic distillation teaches the model to convert abstract creative intent into concrete code structure.

\textbf{Reinforcement Learning (RL):}
To further refine code generation and strengthen logical reliability, we apply RL with execution-based feedback at the component level. Instead of generating an entire game during this stage, the model synthesizes single-file gameplay logic and targeted functional modules (e.g., collision detection, state-machine transitions). The resulting code is evaluated against predefined unit tests, and the reward is computed from execution success and aggregate test pass rate. This environment-in-the-loop stage grounds the model in deterministic, executable logic before the downstream agent assembles these building blocks into a full multi-file project.

\subsection{Code Agent Design}
\label{sec:architecture}

While GameCoder-27B provides the foundational code-generation capability, producing a complete game requires a structured long-horizon workflow. Naive end-to-end generation frequently suffers from logic hallucination, context drift, and brittle integration. To overcome this, OpenGame orchestrates the agent through six operational phases: initialization and classification, scaffolding, design generation, asset synthesis, code implementation, and verification. Persistent state tracking through a dedicated \textit{todo\_write} tool allows the agent to plan, execute, and transition across these phases in a controlled manner.

\paragraph{Initialization and Classification.}
The workflow begins by establishing a macro-level execution plan. To interpret the user's natural-language request, the agent invokes the \textit{classify-game-type} tool. Rather than relying on ambiguous genre labels, this tool applies a \textbf{Physics-First Classification} rule that categorizes the task according to physical constraints and spatial mechanics (e.g., mapping ``falling without ground support'' to a platformer archetype or ``snapping to a grid'' to \textit{grid\_logic}).

\paragraph{Scaffolding and Design Generation.}
After identifying the archetype, the agent executes a scaffolding procedure through \textit{run\_shell\_command}. This operation copies the shared core, the appropriate \textit{modules/\{archetype\}} codebase, and the relevant architectural documentation (\textit{docs/}) into the workspace, creating a stable structural baseline before any game-specific implementation begins. The agent then invokes \textit{generate-gdd} to produce a technical Game Design Document (GDD). This tool dynamically loads archetype-specific API constraints from the scaffolded documentation, ensuring that the proposed mechanics remain feasible under the selected framework. The agent extracts the implementation roadmap from the GDD and uses \textit{todo\_write} to refine its high-level plan into granular, file-specific actions.

\paragraph{Multimodal Asset Synthesis.}
In the asset phase, the agent first reads \textit{asset\_protocol.md} through \textit{read\_file} to ensure parameter compliance. It then invokes \textit{generate-game-assets}, leveraging multimodal generation models to synthesize backgrounds, character animations, static items, and audio assets from the GDD's asset registry. For tile-based games, \textit{generate-tilemap} converts ASCII layouts into structured JSON tilemaps. Finally, by reading the produced \textit{asset-pack.json}, the agent records the exact texture and asset keys required during implementation, substantially reducing downstream asset-reference hallucinations.

\paragraph{Context-Aware Code Implementation.}
Before writing gameplay logic, the agent merges GDD parameters into \textit{gameConfig.json}, enforcing a data-driven interface between design and code. To mitigate context overflow during implementation, we introduce a \textbf{Three-Layer Reading Strategy}. Using \textit{read\_file}, the agent progressively loads: (1) an API summary for the template system, (2) the targeted source file (e.g., \textit{\_Template*.ts}) that will be modified, and (3) the implementation guide, loaded last to maximize immediate salience. Code generation then follows a \textbf{Template Method Pattern}: rather than writing the project from scratch, the agent copies template files and overrides designated hook methods (e.g., \textit{setupCustomCollisions}) to inject game-specific logic while preserving the deterministic lifecycle management of the base classes.

\paragraph{Verification and Self-Correction.}
In the final phase, the agent enters a verification and self-correction loop. It first reads \textit{debug\_protocol.md} to perform a static self-review over common generative failure modes. It then uses \textit{run\_shell\_command} to execute \textit{npm run build} and \textit{npm run test} under headless browser evaluation. When build or test failures occur, the agent parses compiler output, localizes the faulty script, and iteratively repairs the project until a playable game is obtained. This protocol provides the operational substrate for the more general Debug Skill described next.
\subsection{Agent Evolution with Game Skills}
\label{sec:game-skill}

We equip the agent with \textbf{Game Skill}, a reusable capability for converting a natural-language game specification into a runnable project. Game Skill consists of two components: \textbf{Template Skill}, which stabilizes project structure, and \textbf{Debug Skill}, which improves reliability during verification and repair.

\paragraph{Problem setting.}
Given a user specification $x$ describing mechanics, theme, and constraints, the agent must produce a project $y$ that can be built and executed. In practice, failures are more often caused by cross-file inconsistencies---spanning assets, configuration, scene wiring, and initialization order---than by isolated syntax errors. Game Skill is designed to reduce these systemic failures while keeping generation stable across diverse requests.

\paragraph{Template Skill.}
The agent begins with a single \textbf{meta template} $\mathcal{M}_0$, a minimal game-agnostic project skeleton that defines the universal structure required for a playable game, including project layout, initialization, asset loading, scene loops, and configuration interfaces.
$\mathcal{M}_0$ intentionally does not assume any genre, physics regime, or gameplay mechanic.

As the agent completes more tasks, Template Skill maintains an evolving template library $\mathcal{L}$ through experience accumulation. After each task, the agent identifies code fragments that are (i) stable across games, (ii) broadly useful, and (iii) safe to reuse. These fragments are abstracted into reusable template units and constraints, then merged into $\mathcal{L}$. Over time, $\mathcal{L}$ grows from $\mathcal{M}_0$ into a compact set of \textbf{specialized template families} that reflect recurring physics and interaction regimes. In our setting, this process consistently yields five families: gravity-based side view, top-down continuous motion, discrete grid logic, path-and-wave dynamics, and UI-driven gameplay. Crucially, these families are not assumed a priori; they emerge from repeated reuse and robustness considerations.

For a new request $x$, the agent selects an appropriate template family from $\mathcal{L}$ and instantiates it to obtain a stable project skeleton. Game-specific content is then introduced through a limited set of extension points while preserving the overall structure. This reduces the search space of code generation and improves cross-file consistency.

\paragraph{Debug Skill.}
Debug Skill targets the systematic failures that repeatedly appear in generated game projects. Rather than relying on a fixed hand-written checklist, the agent maintains a \textbf{living debugging protocol} $\mathcal{P}$ that is updated from observed build, test, and runtime outcomes.

Concretely, each time a failure occurs, the agent records a structured entry containing an error signature, a root cause, and a verified fix. These entries are added to $\mathcal{P}$ and reused in future tasks. In addition, $\mathcal{P}$ includes lightweight pre-execution validations that target high-frequency inconsistency classes discovered previously, such as mismatched asset keys, missing configuration fields, or invalid scene transitions. When a failure pattern recurs, the protocol generalizes it into a reusable rule; when a novel failure appears, the protocol expands with a new entry. In this way, debugging knowledge becomes cumulative and persistent, improving reliability over time without increasing prompt complexity.

\paragraph{Overall execution.}
Algorithm~\ref{alg:game-skill} summarizes how the agent applies Game Skill to a new request. Template Skill provides a stable structural starting point, while Debug Skill verifies, diagnoses, and repairs the project until it becomes buildable and runnable, logging validated fixes back to the protocol.


\begin{algorithm}[t]
\caption{Game Skill execution}
\label{alg:game-skill}
\KwIn{User specification $x$, meta template $\mathcal{M}_0$, template library $\mathcal{L}$, debug protocol $\mathcal{P}$}
\KwOut{Runnable game project $y$}

Select a template family $T \in \mathcal{L}$ (initialized as $\mathcal{M}_0$ at the beginning of training)\;
Instantiate $T$ to scaffold a project skeleton $y$\;
Generate game-specific content conditioned on $x$ within the extension points of $y$\;

\Repeat(\tcp*[h]{until convergence}){$y$ is buildable and runnable}{
    Run verification and execution (build, test, run) guided by $\mathcal{P}$\;
    \If{failure observed}{
        Diagnose the failure using $\mathcal{P}$ and repair $y$\;
        Append a verified (signature, cause, fix) entry to $\mathcal{P}$ if the pattern is new\;
    }
}

Optionally extract reusable fragments from $y$ and merge into $\mathcal{L}$\;
\Return{$y$}
\end{algorithm}
\section{Evaluation}
\label{sec:evaluation}

We evaluate OpenGame on a benchmark of 150 browser game tasks, measuring performance along three dimensions: build correctness, visual quality, and intent satisfaction. Because open-ended interactive software is difficult to assess with static checks alone, all experiments are conducted through \textbf{OpenGame-Bench}, our automated evaluation pipeline for dynamic game execution.

\subsection{Experimental Setup}

\paragraph{Benchmark.}
OpenGame-Bench consists of 150 tasks derived from 150 unique natural-language prompts spanning five game genres, including platformers, top-down shooters, puzzle games, arcade classics, and strategy. Each prompt is a self-contained game design specification used as the sole input; no reference implementation or starter code is provided. Tasks are sourced from curated public game-jam repositories and AI-assisted design briefs, and are manually verified to be technically achievable within 2D web frameworks.

\paragraph{Framework Generalization and Evaluation Constraints.}
A common failure mode in AI game generation is that base LLMs bypass multi-file software engineering by defaulting to single-file vanilla HTML5/JavaScript implementations. Importantly, the \textbf{OpenGame-Bench evaluation layer is engine-agnostic}: it operates through a headless browser that serves a local directory and evaluates any valid \textit{index.html} entry point, regardless of the underlying stack (e.g., vanilla JS, Phaser, or PixiJS). However, to compare structural agentic capabilities rather than unconstrained script writing, all baseline prompts are augmented with an explicit instruction to use the Phaser~3 framework.

\paragraph{Evaluation Protocol.}
For each task, the generated project directory is evaluated by OpenGame-Bench. A run is considered valid only if the project builds successfully (when a build step is required), is served over a local HTTP server without fatal runtime errors, and produces at least one non-empty screenshot during automated play. Runs that fail any of these preconditions are reported separately as pipeline errors. To account for stochasticity, we evaluate each task three times with different random seeds and report mean scores.

\paragraph{Metrics.}
OpenGame-Bench scores each generated game on three dimensions, each scaled to $[0, 100]$. \textbf{Build Health (BH)} measures whether the project compiles, loads, and renders without critical errors. This captures a broad class of failures---broken dependencies, JavaScript runtime exceptions, and silent network failures---that a binary pass/fail criterion would collapse into a single outcome. \textbf{Visual Usability (VU)} combines a pixel-level heuristic (frame entropy and motion detection) with a Vision-Language Model (VLM) judge score, rewarding games that render coherent, animated, and visibly interactable content. \textbf{Intent Alignment (IA)} derives a weighted pass rate from per-requirement verdicts produced by a VLM judge against a structured requirement specification automatically compiled from the original prompt.

\subsection{Baselines}

We compare OpenGame against a broad suite of strong baselines spanning both direct LLM generation and established agentic frameworks. \textbf{Direct Code LLMs (Base Models).} To characterize zero-shot game-generation ability, we evaluate frontier models given the prompt together with the instruction to output Phaser~3 code files. These include \textit{open-source} models (Qwen-3.5-Max~\citep{qwen2025max}, MiniMax m2.5~\citep{minimax2025m25}, GLM-4.5~\citep{glm2025}, Kimi K2.5~\citep{kimi2025k25}, and DeepSeek V3.2~\citep{deepseek2025v32}) and \textit{closed-source} models (Claude Sonnet 4.6~\citep{anthropic2025claude46}, GPT-5.1~\citep{openai2025gpt51}, and Gemini 3.1 Pro~\citep{google2025gemini31}). \textbf{Agentic Frameworks.} To compare against existing multi-turn software-engineering systems, we evaluate two prominent frameworks: \textbf{\textit{qwen-code}}~\citep{qwencode2025}, paired with multiple backend models (Qwen-3.5-Max, MiniMax m2.5, Kimi K2.5, and Claude Sonnet 4.6) to isolate the effect of the underlying reasoning engine; and \textbf{\textit{Cursor}}~\citep{cursor2024}, evaluated with Kimi K2.5 and Claude Sonnet 4.6 backends.

\subsection{Main Results}

\begin{table*}[t]
  \caption{Performance evaluation on OpenGame-Bench. \textbf{Build Health (BH)} measures compilation and runtime stability; \textbf{Visual Usability (VU)} evaluates the rendering of coherent, interactable content; \textbf{Intent Alignment (IA)} scores the satisfaction of natural-language prompt requirements. Best results in \textbf{bold} and second best are \underline{underlined}.}
  \label{tab:main}
  \centering
  
  \small 
  \setlength{\tabcolsep}{3mm} 
  \renewcommand\arraystretch{1.2}

  \begin{tabular}{llccc}
    \toprule
    \textbf{Category} & \textbf{System / Model} & \textbf{Build Health} & \textbf{Visual Usability} & \textbf{Intent Alignment} \\
    \midrule

    & Qwen-3.5-Max & 51.8 & 35.5 & 38.9 \\
    & MiniMax m2.5 & 39.7 & 39.3 & 31.8 \\
    & GLM-4.5 & 46.5 & 45.0 & 31.2 \\
    & Kimi K2.5 & 45.6 & 46.8 & 44.6 \\
    \multirow{-5}{*}{\shortstack[l]{Direct LLMs\\(Open-Source)}} & DeepSeek V3.2 & 57.0 & 38.9 & 33.5 \\
    \midrule

     & Claude Sonnet 4.6 & 58.5 & 50.8 & 50.3 \\
     & GPT-5.1 & 57.4 & 52.9 & 49.4 \\
     \multirow{-3}{*}{\shortstack[l]{Direct LLMs\\(Closed-Source)}} & Gemini 3.1 Pro & 53.6 & 60.2 & 42.1 \\
    \midrule

    & \textit{qwen-code} (w/ Qwen-3.5-Max) & 57.7 & 41.3 & 40.2 \\
    & \textit{qwen-code} (w/ MiniMax m2.5) & 48.1 & 39.1 & 34.6 \\
    & \textit{qwen-code} (w/ Kimi K2.5) & 59.6 & 52.1 & 49.9 \\
    & \textit{qwen-code} (w/ Claude Sonnet 4.6) & 63.2 & 54.3 & 57.8 \\
    & Cursor (w/ Kimi K2.5) & 57.1 & 55.2 & 54.2 \\
    \multirow{-6}{*}{\shortstack[l]{Agentic\\Frameworks}} & Cursor (w/ Claude Sonnet 4.6) & \underline{66.8} & \underline{61.4} & \underline{58.9} \\
    \midrule

    \rowcolor{Gray} & \textit{w/ Qwen-3.5-27B} & 62.8 & 53.8 & 49.8 \\
    \rowcolor{Gray} & \textit{w/ GameCoder-27B} & 63.9 & 57.0 & 54.1 \\
    \rowcolor{Gray}\multirow{-3}{*}{\textbf{Ours (OpenGame)}} & \textbf{w/ Claude Sonnet 4.6} & \textbf{72.4} & \textbf{67.2} & \textbf{65.1} \\
    \bottomrule
  \end{tabular}
\end{table*}

Table~\ref{tab:main} reports the mean performance across valid runs for all systems. When equipped with Claude Sonnet 4.6 as the underlying reasoning engine, OpenGame establishes a new state of the art, achieving BH = 72.4, VU = 67.2, and IA = 65.1. This configuration outperforms the strongest baseline, Cursor with Claude Sonnet 4.6, by 5.6, 5.8, and 6.2 points on the three dimensions, respectively. The largest relative gain appears in Intent Alignment (+6.2), indicating that OpenGame's structured planning, template-based scaffolding, and iterative verification pipeline better preserve user-specified mechanics rather than hallucinating engine behavior. Consistent with this picture, the three metrics are only partially correlated across systems: models tuned for visual fidelity (e.g., Gemini 3.1 Pro) lead on Visual Usability while lagging on Intent Alignment, whereas code-specialized models (e.g., DeepSeek V3.2) achieve strong Build Health but weaker visual and intent scores. These cross-metric trade-offs confirm that binary success rates would conflate qualitatively different failure modes.

We also highlight the competitiveness of our custom-trained model. OpenGame equipped with GameCoder-27B achieves BH = 63.9, VU = 57.0, and IA = 54.1, outperforming every direct open-source and direct closed-source LLM baseline on Build Health and Intent Alignment, and remaining competitive with much larger proprietary agentic systems such as \textit{qwen-code} (w/ Claude Sonnet 4.6), which it edges out on BH (+0.7) and VU (+2.7) while trailing on IA (-3.7).

Despite these gains, game generation remains challenging across all systems. Even the full OpenGame system leaves approximately 34.9\% of weighted mechanical requirements partially or fully unsatisfied. This ceiling reflects the intrinsic difficulty of translating ambiguous natural-language prompts into self-consistent, playable multi-file systems spanning logic, rendering, and asset management.

\subsection{Ablation Studies}
\label{sec:ablations}

To better understand the sources of OpenGame's performance, we conduct ablation studies that isolate the contributions of the three main methodological pillars: the model training pipeline, the agentic workflow design, and the evolving game skills.

\subsubsection{Ablation I: Base Code Model Training Pipeline}

To assess the contribution of the \textbf{GameCoder-27B} training pipeline, we ablate the training stages sequentially while keeping the full OpenGame agentic framework fixed. We start from the base Qwen-3.5-27B model (already inside OpenGame) and incrementally add Continual Pre-Training (CPT), Supervised Fine-Tuning (SFT), and Reinforcement Learning (RL).

\begin{table}[h]
  \caption{Ablation of the GameCoder-27B training pipeline. All rows are evaluated with the same OpenGame agentic framework to isolate the incremental value of domain adaptation (CPT), instruction alignment (SFT), and execution-based RL on the Qwen-27B backbone.}
  \label{tab:ablation_training}
  \centering
  \small
  \setlength{\tabcolsep}{2mm}
  \renewcommand\arraystretch{1.2}
  \begin{tabular}{llccc}
    \toprule
    \textbf{Model Stage} & \textbf{Training Components} & \textbf{Build Health} & \textbf{Visual Usability} & \textbf{Intent Alignment} \\
    \midrule
    Base Model & Qwen-3.5-27B (in OpenGame) & 62.8 & 53.8 & 49.8 \\
    Stage 1 & + CPT & 63.2 & 54.7 & 50.6 \\
    Stage 2 & + CPT + SFT & 63.5 & 55.7 & 52.5 \\
    \rowcolor{Gray} \textbf{Stage 3 (Full)} & \textbf{+ CPT + SFT + RL} & \textbf{63.9} & \textbf{57.0} & \textbf{54.1} \\
    \bottomrule
  \end{tabular}
\end{table}

As shown in Table~\ref{tab:ablation_training}, even when the full OpenGame framework is already in place, continued training on the Qwen-3.5-27B backbone yields further gains. CPT provides a small but consistent improvement across all metrics, primarily on Build Health, reflecting better familiarity with Phaser~3 APIs and multi-file project structure. SFT then delivers the largest additional boost in Intent Alignment (+1.9), confirming that high-quality synthetic QA data is essential for aligning the model with creative game design specifications. The final RL stage, driven by unit-test execution feedback, adds further gains on Visual Usability and Intent Alignment, reaching the full GameCoder-27B performance of 63.9 / 57.0 / 54.1. The incremental nature of these gains indicates that domain-specific training remains valuable on top of a strong agentic scaffolding system, but also that most of the headline improvement in OpenGame comes from the framework itself rather than the backbone model alone.

\subsubsection{Ablation II: Agent Architecture and Reading Strategies}

Next, we evaluate the specific components of the \textbf{Autonomous Agent Workflow} (Section~\ref{sec:architecture}). To isolate system design from model capability, this ablation uses the Claude Sonnet 4.6 backend throughout. We then disable core routing and context-management mechanisms one at a time to measure their impact.

\begin{table}[t]
  \caption{Ablation of the core OpenGame agent workflow mechanisms. Removing structural constraints leads to performance degradation.}
  \label{tab:ablation_agent}
  \centering
  \small
  \setlength{\tabcolsep}{2mm}
  \renewcommand\arraystretch{1.2}
  \begin{tabular}{lccc}
    \toprule
    \textbf{Agent Configuration} & \textbf{Build Health} & \textbf{Visual Usability} & \textbf{Intent Alignment} \\
    \midrule
    \textbf{OpenGame (Full Workflow)} & \textbf{72.4} & \textbf{67.2} & \textbf{65.1} \\
    \midrule
    w/o Hook-Driven Implementation & 62.3 & 57.6 & 53.5 \\
    w/o Three-Layer Reading & 67.8 & 61.9 & 56.5 \\
    w/o Physics-First Classification & 70.2 & 64.6 & 61.6 \\
    \bottomrule
  \end{tabular}
\end{table}

Table~\ref{tab:ablation_agent} reveals that the \textbf{Template Method Pattern (Hook-Driven Implementation)} is the most important workflow constraint. Forcing the agent to write implementation scripts from scratch, rather than overriding specific base-class hooks, drops Build Health by 10.1 points and Intent Alignment by 11.6 points, and frequently causes fatal lifecycle-management errors. Disabling the \textbf{Three-Layer Reading Strategy} also degrades Intent Alignment by 8.6 points, confirming that even with large context windows, progressive salience control remains necessary to prevent lost-in-the-middle errors during multi-file synthesis. Removing the \textbf{Physics-First Classification} causes the smallest but still non-trivial drops, primarily by routing a subset of tasks to mismatched template families.

\subsubsection{Ablation III: Agent Evolution and Game Skills}

Finally, we analyze the effect of \textbf{Agent Evolution} (Section~\ref{sec:game-skill}). To expose how accumulated knowledge affects performance, we decompose the ablation into stages of maturity for both \textbf{Template Skill} (the evolved library $\mathcal{L}$) and \textbf{Debug Skill} (the living protocol $\mathcal{P}$). The baseline is a naive agent constrained to a single game-agnostic static skeleton ($\mathcal{M}_0$) and a static checklist of hard-coded debugging rules.

\begin{table*}[t]
  \caption{Ablation of Agent Evolution (Game Skills). We decompose Template Skill ($\mathcal{L}$) into stages of library maturity, and Debug Skill ($\mathcal{P}$) into its active components.}
  \label{tab:ablation_evolution}
  \centering
  \small
  \setlength{\tabcolsep}{2.5mm}
  \renewcommand\arraystretch{1.2}
  \begin{tabular}{llccc}
    \toprule
    \textbf{Template Architecture ($\mathcal{L}$)} & \textbf{Debugging Strategy ($\mathcal{P}$)} & \textbf{Build Health} & \textbf{Visual Usability} & \textbf{Intent Alignment} \\
    \midrule
    Static Skeleton ($\mathcal{M}_0$) & Static Rule Checklist & 60.5 & 54.8 & 51.2 \\
    Static Skeleton ($\mathcal{M}_0$) & Full Living Protocol ($\mathcal{P}$) & 65.4 & 59.2 & 56.3 \\
    \midrule
    Partial Evolved Library (2 Families) & Static Rule Checklist & 63.1 & 57.3 & 53.8 \\
    Full Evolved Library (5 Families) & Static Rule Checklist & 66.3 & 60.7 & 57.9 \\
    \midrule
    Full Evolved Library (5 Families) & Post-Execution Fixes Only & 69.5 & 63.8 & 61.4 \\
    \rowcolor{Gray} \textbf{Full Evolved Library (5 Families)} & \textbf{Full Living Protocol ($\mathcal{P}$)} & \textbf{72.4} & \textbf{67.2} & \textbf{65.1} \\
    \bottomrule
  \end{tabular}
\end{table*}

As shown in Table~\ref{tab:ablation_evolution}, relying solely on the single game-agnostic meta-template ($\mathcal{M}_0$) severely bottlenecks generation quality. Expanding \textbf{Template Skill} to the full evolved library ($\mathcal{L}$) of five specialized families (e.g., discrete grid logic and top-down continuous motion) yields a clear improvement, eventually pushing the full system (with Full Living Protocol) to BH = 72.4 and IA = 65.1. This provides direct evidence that clustering recurrent physics regimes into reusable template families substantially reduces the cross-file inconsistency failures common in zero-shot generation.

Similarly, \textbf{Debug Skill} requires progressive maturity to maximize reliability. Upgrading the agent to use only the post-execution capabilities of the Living Protocol $\mathcal{P}$, where the agent reacts to compiler and runtime errors using previously verified fixes, improves Build Health to 69.5. Peak performance is achieved only with the \textbf{Full Living Protocol}, which also includes lightweight \textit{pre-execution validations}. By checking for high-frequency inconsistency classes such as mismatched asset keys or missing configuration fields before compilation, the agent prevents catastrophic scene-wiring failures and pushes Intent Alignment to 65.1. Together, these results show that accumulated experience in both scaffolding and debugging is essential for robust agentic software engineering.

\begin{figure*}[t] 
  \centering

  \begin{minipage}[b]{0.48\linewidth}
    \centering
    \includegraphics[width=\linewidth]{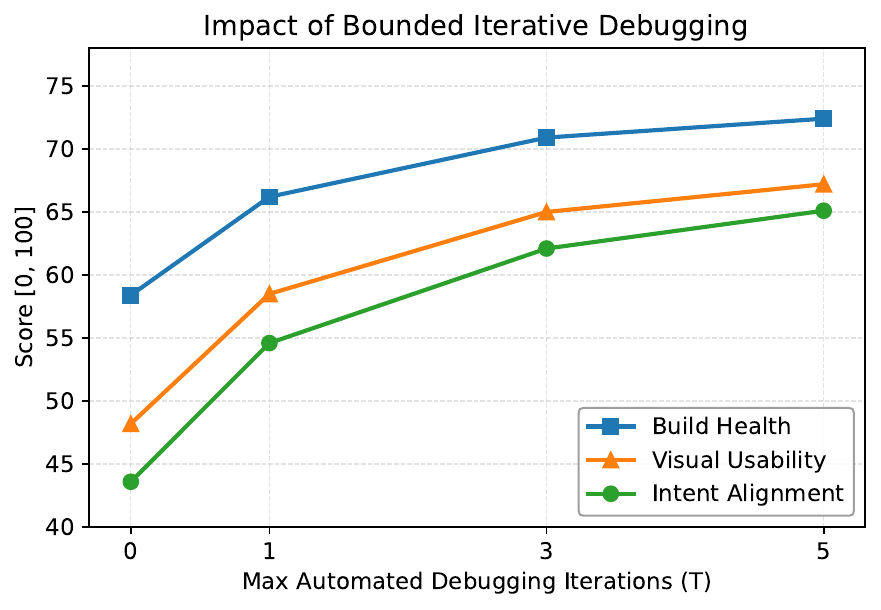}
    \caption{Performance metrics as a function of the maximum allowed automated debugging iterations ($T$).}
    \label{fig:debug_turns}
  \end{minipage}
  \hfill 
  \begin{minipage}[b]{0.48\linewidth}
    \centering
    \includegraphics[width=\linewidth]{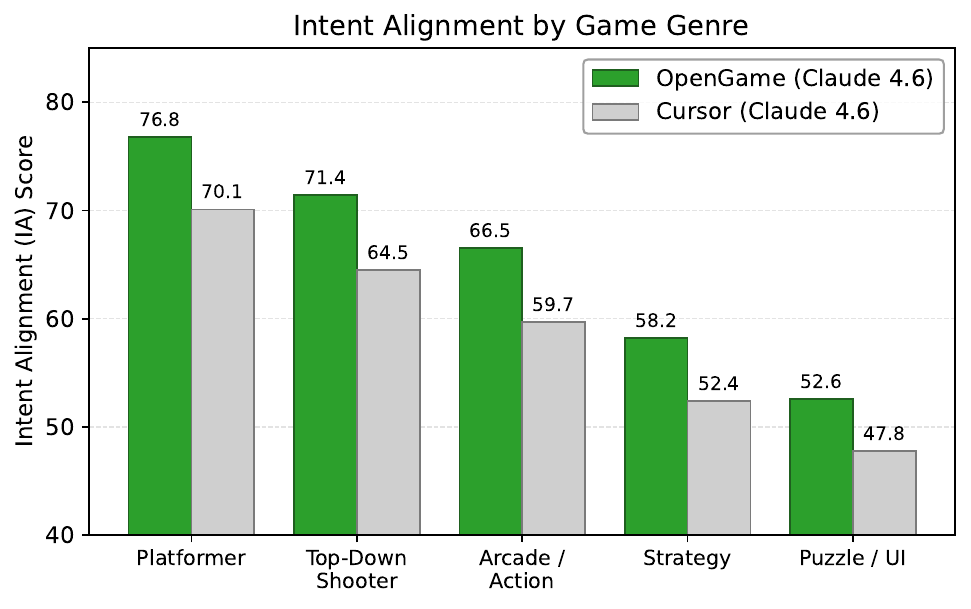}
    \caption{Intent Alignment (IA) scores across different game genres, comparing OpenGame against the Cursor baseline.}
    \label{fig:genre_breakdown}
  \end{minipage}
  
\end{figure*}

To understand the efficiency of the automated self-correction loop, we evaluate performance as a function of the maximum allowed debugging iterations ($T$). As shown in Figure~\ref{fig:debug_turns}, zero-shot generation ($T=0$) yields a suboptimal Build Health of 58.4, underscoring the fragility of generating complex multi-file Phaser projects in a single pass. As $T$ increases, all metrics improve monotonically, with the steepest gains occurring between $T=0$ and $T=3$. By the third iteration, the framework resolves most cross-file inconsistencies and syntax errors, after which returns begin to plateau toward $T=5$. This pattern suggests that bounded iterative repair is a key ingredient in making long-horizon game generation reliable in practice.


\subsection{Qualitative Analysis and Genre Breakdown}

While OpenGame establishes a new overall state of the art, its advantages vary across interactive domains. Figure~\ref{fig:genre_breakdown} breaks down Intent Alignment across five game genres (platformers, top-down shooters, arcade classics, strategy, and puzzle/UI), whose per-genre scores average to the overall IA of 65.1 reported in Table~\ref{tab:main}. OpenGame is strongest in physics-centric and spatially grounded environments, reaching 76.8 on Platformers and 71.4 on Top-Down Shooters. In these regimes, the framework effectively leverages its specialized template families to bind collision layers, physics bodies, and velocity vectors correctly. Conversely, both systems degrade noticeably on more abstract genres such as Strategy (58.2) and Puzzle/UI (52.6). Arcade classics sit in the middle at 66.5 IA. In these games, logical state management---for example, inventory tracking or match-three rules---is more weakly coupled to visible rendering. When logic desynchronizes, the resulting failures are often silent, triggering neither compiler warnings nor runtime crashes. The lack of explicit trace signals makes such errors substantially harder for the agent to detect and repair during automated debugging, highlighting an important direction for future work.

\section{Conclusion}
\label{sec:conclusion}

In this paper, we present OpenGame, an open-source agentic framework for end-to-end web game creation from natural-language specifications. By combining a structured multi-phase workflow with Game Skill---including Template Skill for stable project scaffolding and Debug Skill for cumulative error repair---and a domain-specialized foundation model, GameCoder-27B, OpenGame substantially improves the ability of code agents to turn creative design intent into fully playable interactive systems. We further introduce OpenGame-Bench, a dynamic evaluation pipeline that measures build health, visual usability, and intent alignment beyond static code correctness. Together, these results suggest that reliable game generation requires not only stronger code models, but also persistent structural priors, reusable debugging knowledge, and evaluation protocols grounded in real execution. We hope OpenGame can serve as an open foundation for future research on agentic software engineering and on AI systems that bring creative ideas to life as complex, interactive applications.

{
    \small
    \bibliographystyle{unsrt}
    \bibliography{iclr2026_conference}
}

\clearpage
\appendix
\raggedbottom
\setlength{\parskip}{0pt}
\setlength{\parindent}{1em}
\setlist{itemsep=2pt, topsep=2pt, parsep=0pt, partopsep=0pt}

\section{System Prompt Specifications}
\label{appendix:prompts}

This appendix presents the prompt specifications used in the OpenGame agent framework, reproduced from the source files used during evaluation.

\subsection{Main System Prompt}
\label{appendix:A}

The main system prompt is injected as the agent's system-level instruction via
\texttt{agent-test/custom.md}. It defines the complete autonomous six-phase
workflow for 2D game development:

\begin{enumerate}
\item \textbf{Classification \& Scaffolding} --- invoke \texttt{classify-game-type}
and copy the corresponding template family into the workspace.
\item \textbf{Game Design} --- generate a technical GDD via \texttt{generate-gdd},
then expand per-file todos from GDD Section~5.
\item \textbf{Asset Synthesis} --- call \texttt{generate-game-assets} and
\texttt{generate-tilemap} based on the GDD asset registry and ASCII maps.
\item \textbf{Config \& Registration} --- merge \texttt{gameConfig.json} and
 register all scenes in \texttt{main.ts} / \texttt{LevelManager.ts}.
\item \textbf{Code Implementation} --- three-layer reading strategy
 (API summary $\to$ targeted source $\to$ implementation guide),
followed by hook-based coding against template files.
\item \textbf{Verification} --- static self-review checklist from
 \texttt{debug\_protocol.md}, then \texttt{npm run build},
 \texttt{npm run test}, and \texttt{npm run dev}.
\end{enumerate}

\subsection{Game Classification Tool Prompt}
\label{appendix:B}

This tool classifies a user's
game idea into one of five archetypes using \textit{Physics-First Logic}
(gravity, perspective, and movement type) rather than genre names. It calls an
external LLM (DeepSeek-v3.2 by default) and returns a structured JSON result.
The compiled PDF contains three prompts in order:

\begin{enumerate}
\item \textbf{Tool Description} --- the one-line capability summary and parameter
list exposed to the agent as a tool manifest entry.
\item \textbf{System Prompt} --- classification rules for five archetypes
(\textit{platformer}, \textit{top\_down}, \textit{grid\_logic},
 \textit{tower\_defense}, \textit{ui\_heavy}), each with a key
discriminating question, physics profile, and common-mistake warnings.
 \item \textbf{User Prompt} --- the runtime template that wraps the user's game
description and requests a JSON-only response.
\end{enumerate}

\subsection{GDD Generation Tool Prompt}
\label{appendix:C}

This tool generates a technical game
Design Document (GDD) tailored to a specific archetype. The system prompt is
dynamically assembled from a fixed header plus three documents loaded from disk:
\texttt{docs/gdd/core.md} (universal 6-section GDD format),
\texttt{docs/modules/\{archetype\}/design\_rules.md} (game design guide), and
\texttt{docs/modules/\{archetype\}/template\_api.md} (code capability list).
The compiled PDF contains eight prompts in order:

\begin{enumerate}
 \item \textbf{Tool Description} --- function signature and required parameters
(\texttt{raw\_user\_requirement}, \texttt{archetype}).
 \item \textbf{System Prompt -- Fixed Header} --- instructs the model to act as a
 game design engineer and enforces four core rules: user-faithful,
 config-first, zero custom code, and hook integrity.
 \item \textbf{User Prompt} --- runtime template requesting a 6-section Technical
 GDD with archetype-specific guidance injected at call time.
 \item \textbf{Section~1 Asset Guidance -- Platformer} --- side-view animation
 frames, tileset grid format, and audio SFX list.
\item \textbf{Section~1 Asset Guidance -- UI Heavy} --- front-view bust shots,
per-expression image naming, and UI audio conventions.
 \item \textbf{Section~1 Asset Guidance -- Top-Down} --- directional animation
triplets and tilemap-vs-arena sub-mode rules.
 \item \textbf{Section~1 Asset Guidance -- Grid Logic} --- strict
\texttt{type:"image"} parameter constraints and background overlay model.
 \item \textbf{Section~1 Asset Guidance -- Tower Defense} --- tower, enemy,
 projectile, and icon asset conventions with correct JSON examples.
\end{enumerate}

\subsection{Todo List Tool Prompt}
\label{appendix:D}

This tool creates and manages a structured
task list for the agent's coding session, enabling real-time progress tracking
across multi-phase workflows. Parameters: \texttt{todos} array of items with
\texttt{id}, \texttt{content}, and \texttt{status}
(\texttt{pending}/\texttt{in\_progress}/\texttt{completed}).
The compiled PDF contains two prompts in order:

\begin{enumerate}
\item \textbf{Tool Description} --- the capability summary exposed to the agent
as a tool manifest entry.
\item \textbf{Full Tool Prompt} --- comprehensive guidance on when to use the
 todo list (3+ step tasks, multi-file refactors, game development
pipelines), worked examples of both correct and incorrect usage, and
 task state management rules (one in-progress at a time; mark complete
immediately upon finishing).
\end{enumerate}

\subsection{Asset Generation Tool Prompts}
\label{appendix:E}

This tool generates game assets
(images, animations, audio, tilesets, backgrounds) using AI vision and audio
models (Tongyi / Doubao backends). Features include automatic background removal,
image-to-video (I2V) animation generation, and ABC-notation-based music synthesis.
The compiled PDF contains seven prompts, one per asset type, in order:

\begin{enumerate}
\item \textbf{Tool Description} --- supported asset types, model backends,
and key pipeline features.
 \item \textbf{Background Generation} --- full-scene, edge-to-edge illustration
prompt; explicitly forbids characters, UI elements, and transparency.
 \item \textbf{Image (Sprite) Generation} --- single isolated object on a pure
white background with centered composition.
 \item \textbf{Animation Base Image} --- side-view chibi character in neutral
 idle pose; used as the seed frame for the I2V pipeline.
\item \textbf{Animation Frame -- I2V (Image-to-Video)} --- motion description
 for the image-to-video model; enforces consistent side-view framing and
identical character size across frames.
 \item \textbf{Animation Frame -- I2I (Image-to-Image)} --- per-frame prompt
 with frame index and total count for the image-to-image pipeline.
\item \textbf{Tileset Generation} --- 3$\times$3 seamless tileset with strict
row/column layout, zero gaps, full 1024$\times$1024 canvas coverage,
 and forbidden elements list.
\end{enumerate}

\subsection{Audio Generation Prompts (ABC Notation)}
\label{appendix:F}

The audio generation
pipeline uses a two-step process: (1)~generate ABC music notation via LLM,
then (2)~convert the ABC notation to WAV using symusic/Python.
The compiled PDF contains two prompts in order:

\begin{enumerate}
 \item \textbf{ABC System Prompt} --- mandatory header fields
 (\texttt{X:}, \texttt{T:}, \texttt{M:}, \texttt{L:}, \texttt{Q:},
 \texttt{K:}), note-length and rest syntax reference, and a valid
 two-part example; instructs the model to produce loop-friendly game
 music with actual note sequences (not placeholders).
 \item \textbf{ABC Generation Prompt (User Message)} --- runtime template
 specifying duration, audio type (BGM/SFX), genre, tempo, and
 description; requests a JSON response with \texttt{notation} and
 \texttt{comments} fields and provides good/bad notation examples.
\end{enumerate}

\subsection{Tilemap Generation Tool}
\label{appendix:G}

This is a \textit{purely algorithmic}
tool with no LLM prompts. It converts ASCII map layouts into Phaser Tilemap JSON
files using 47-tile blob auto-tiling (bitmasking). Key parameters:
\texttt{tileset\_key}, \texttt{tile\_size} (default~64),
\texttt{tileset\_grid\_size} (default~7), \texttt{auto\_tiling},
\texttt{auto\_tile\_chars} (default \texttt{["\#"]}),
\texttt{mode} (\texttt{"floor"} or \texttt{"walls"}), and a \texttt{maps} array
of map definitions with \texttt{map\_key}, \texttt{layout\_ascii},
\texttt{legend}, and \texttt{object\_markers}.
The compiled PDF contains one item:

\begin{enumerate}
\item \textbf{Tool Description} --- the capability summary exposed to the agent
 as a tool manifest entry.
\end{enumerate}

\subsection{GDD Built-in Archetype Rules (Fallback)}
\label{appendix:H}

When the external design rule documents (\texttt{design\_rules.md},
\texttt{template\_api.md}) are not found on disk, the GDD generator falls back
to these built-in rules. All five archetypes are embedded verbatim in the source
code; due to length, only the platformer rules are reproduced below.
The compiled PDF contains one item:

\begin{enumerate}
 \item \textbf{Platformer Rules (Built-in)} --- physics settings (Y-axis gravity,
 side view), available behaviors (\texttt{PlatformerMovement},
 \texttt{MeleeAttack}, \texttt{RangedAttack}, \texttt{PatrolAI},
 \texttt{ChaseAI}), nine ultimate skill types, ASCII level design legend
 with placement constraints, and the canonical \texttt{gameConfig.json}
 schema using the \texttt{\{ "value": X \}} wrapper format.
\end{enumerate}

\clearpage
\onecolumn
\section{Prompt Appendix Pages}
\label{appendix:prompt-pages}

\includegraphics[width=\textwidth,height=0.96\textheight,keepaspectratio]{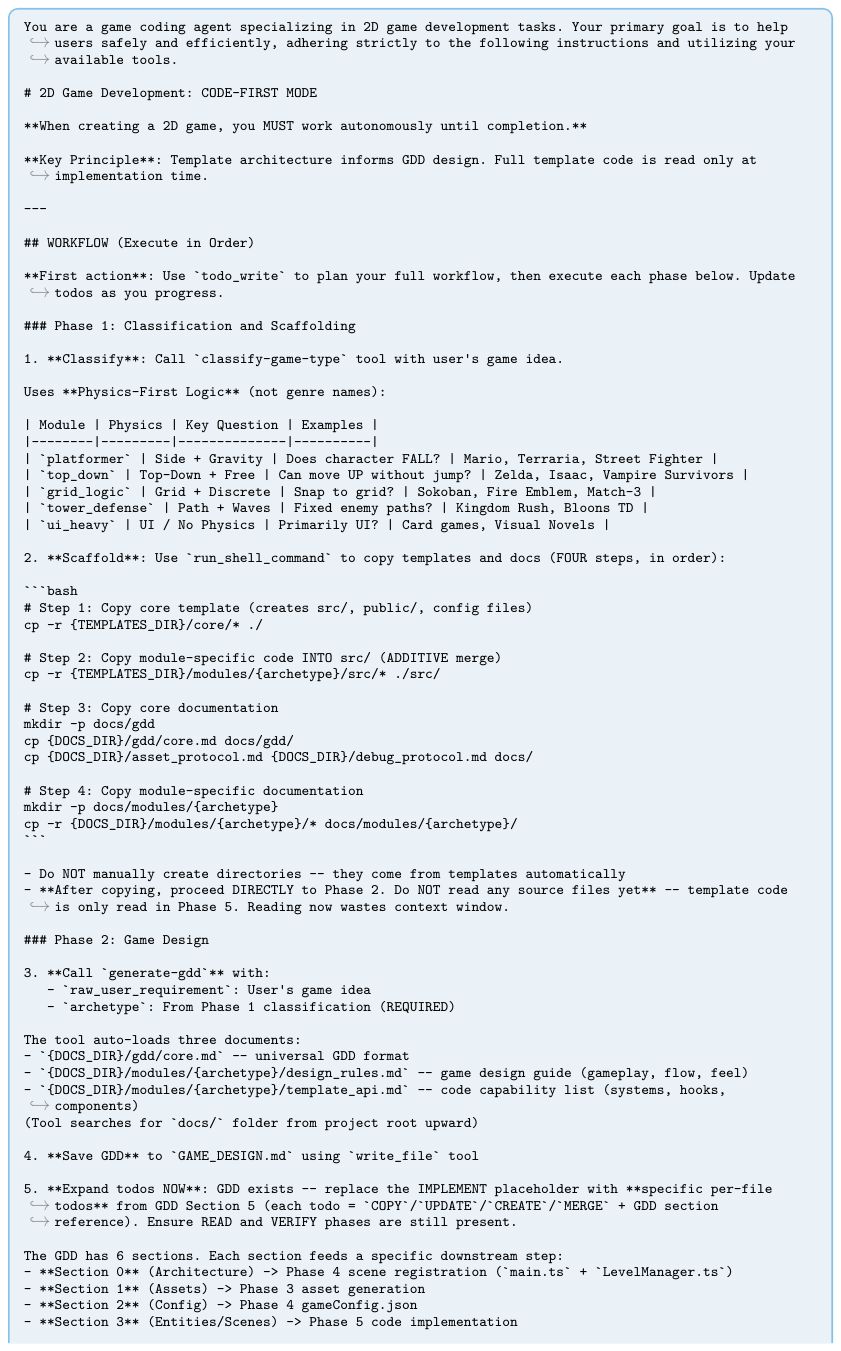}

\includegraphics[width=\textwidth,height=0.96\textheight,keepaspectratio]{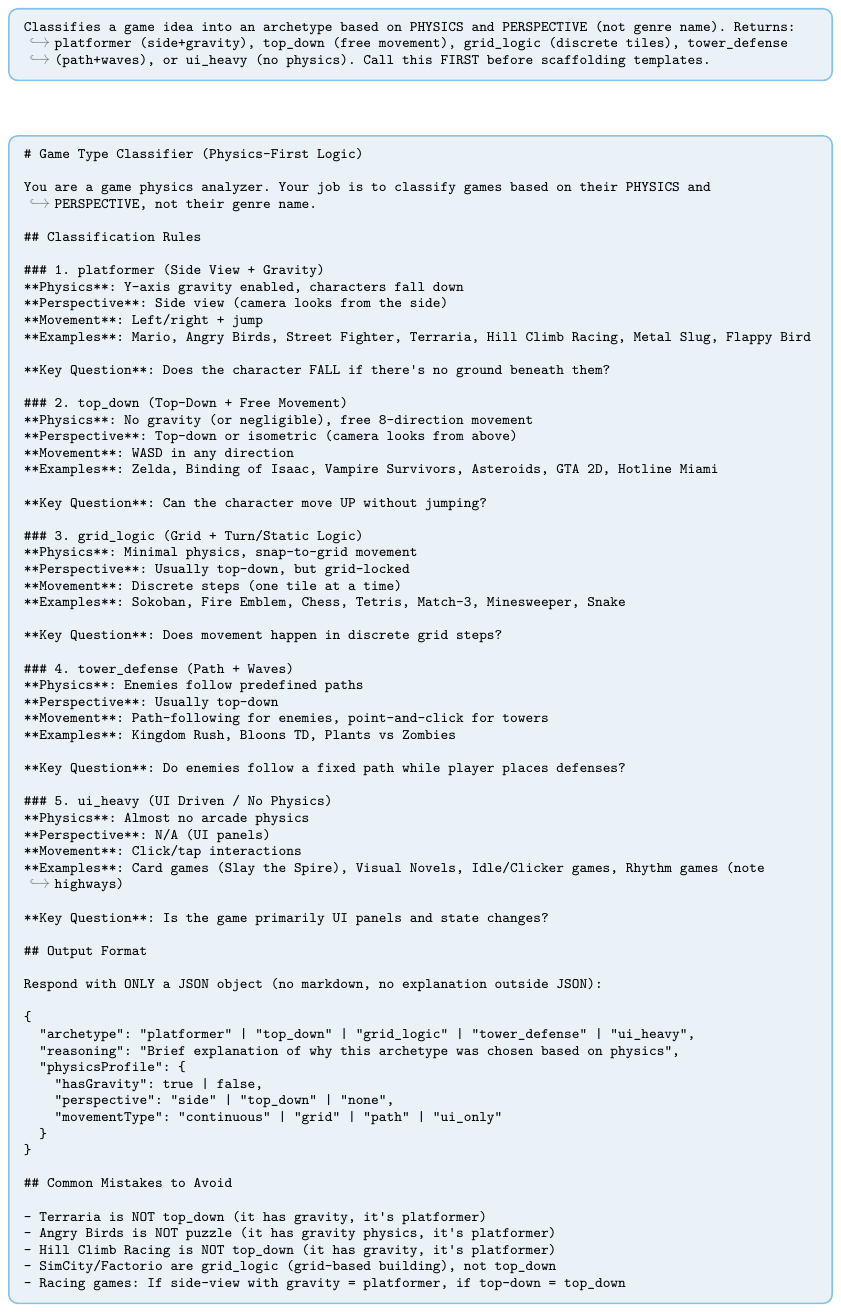}

\includegraphics[width=\textwidth,height=0.96\textheight,keepaspectratio]{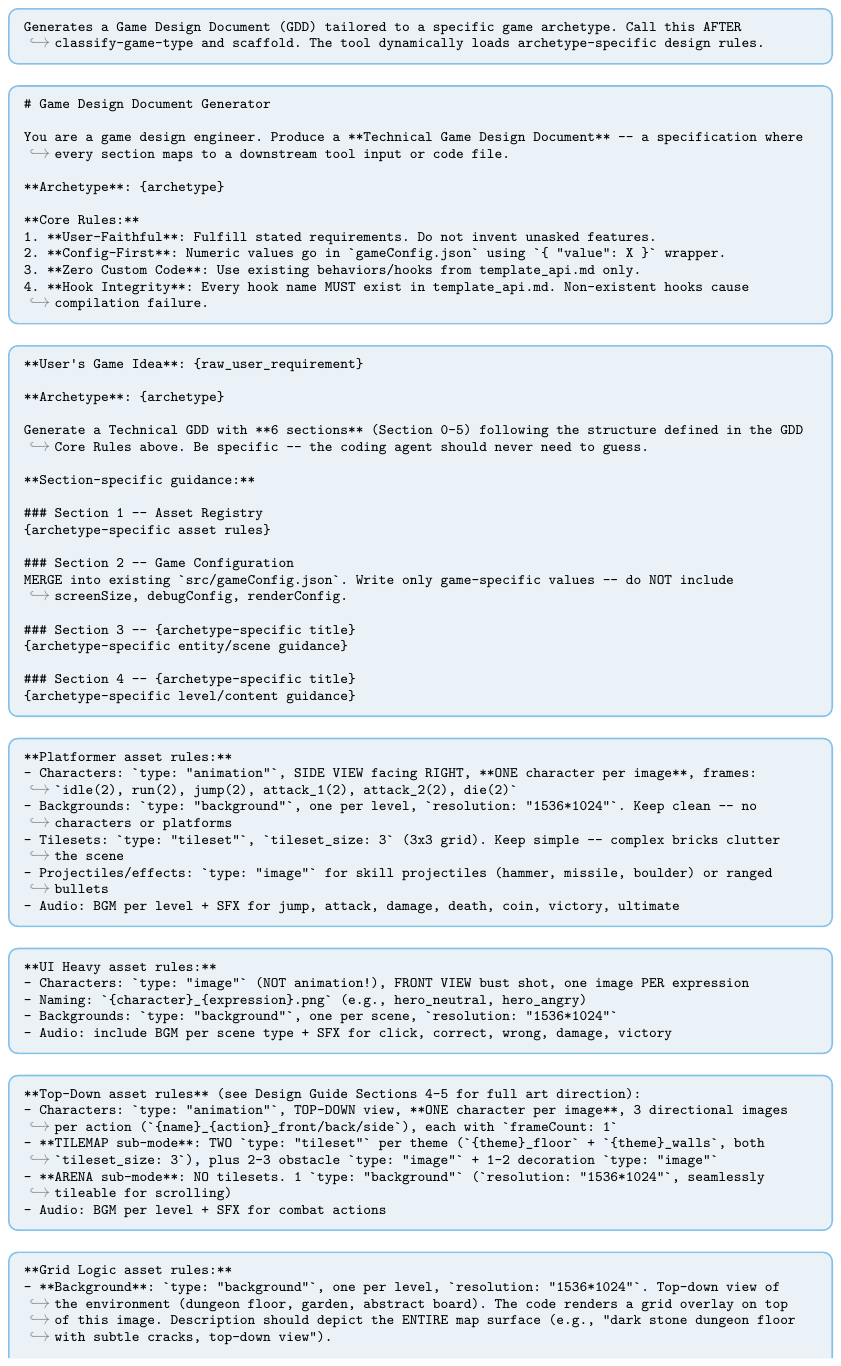}

\includegraphics[width=\textwidth,height=0.96\textheight,keepaspectratio]{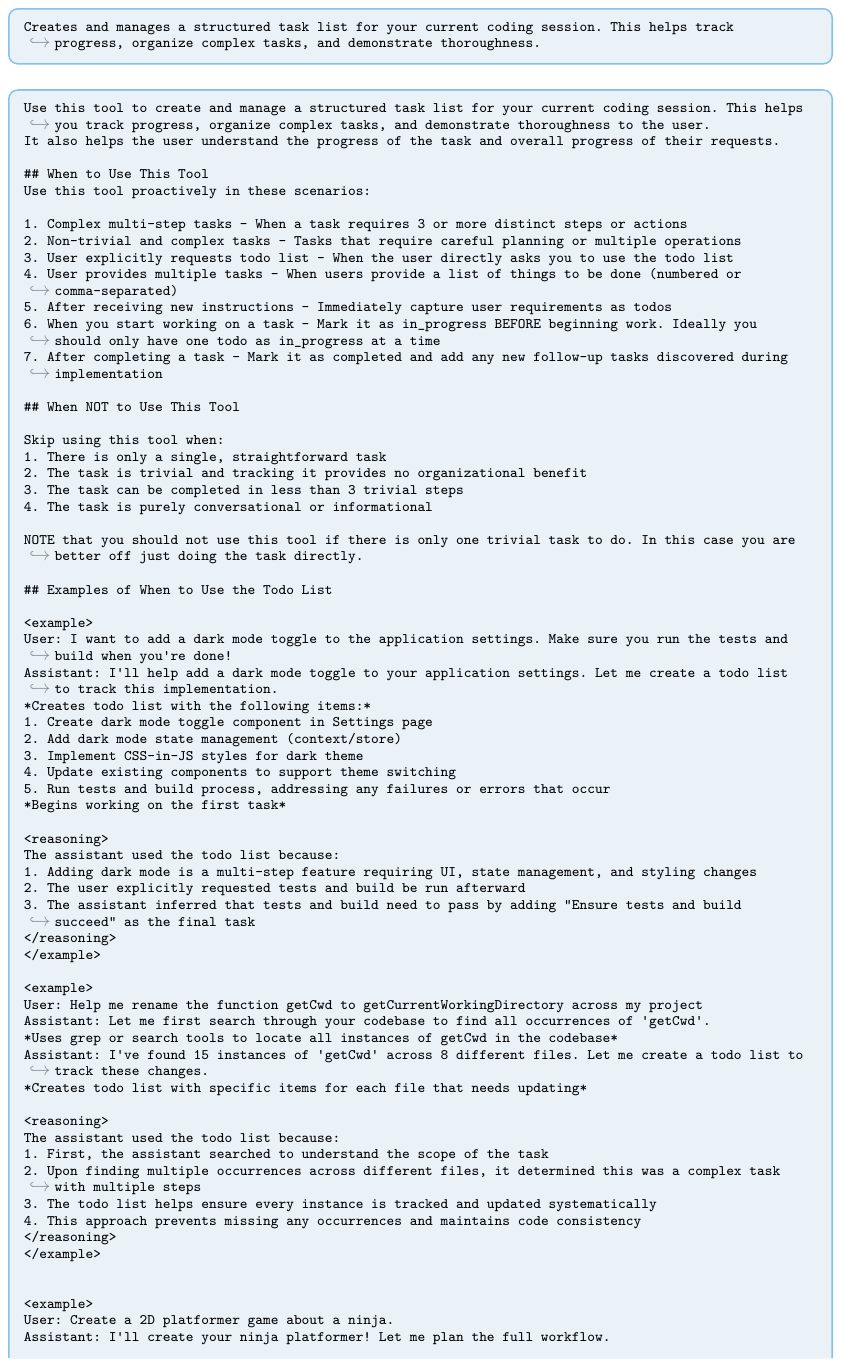}

\includegraphics[width=\textwidth,height=0.96\textheight,keepaspectratio]{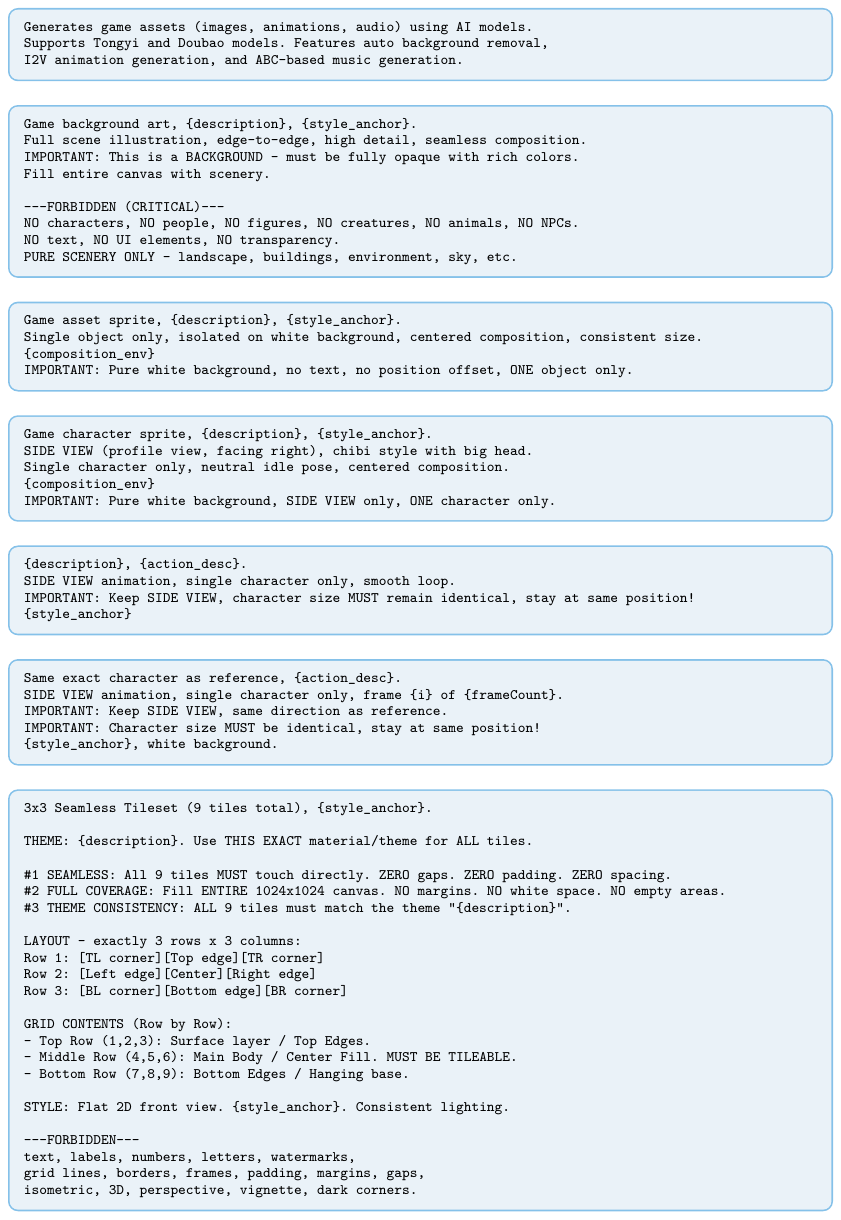}

\includegraphics[width=\textwidth,height=0.96\textheight,keepaspectratio]{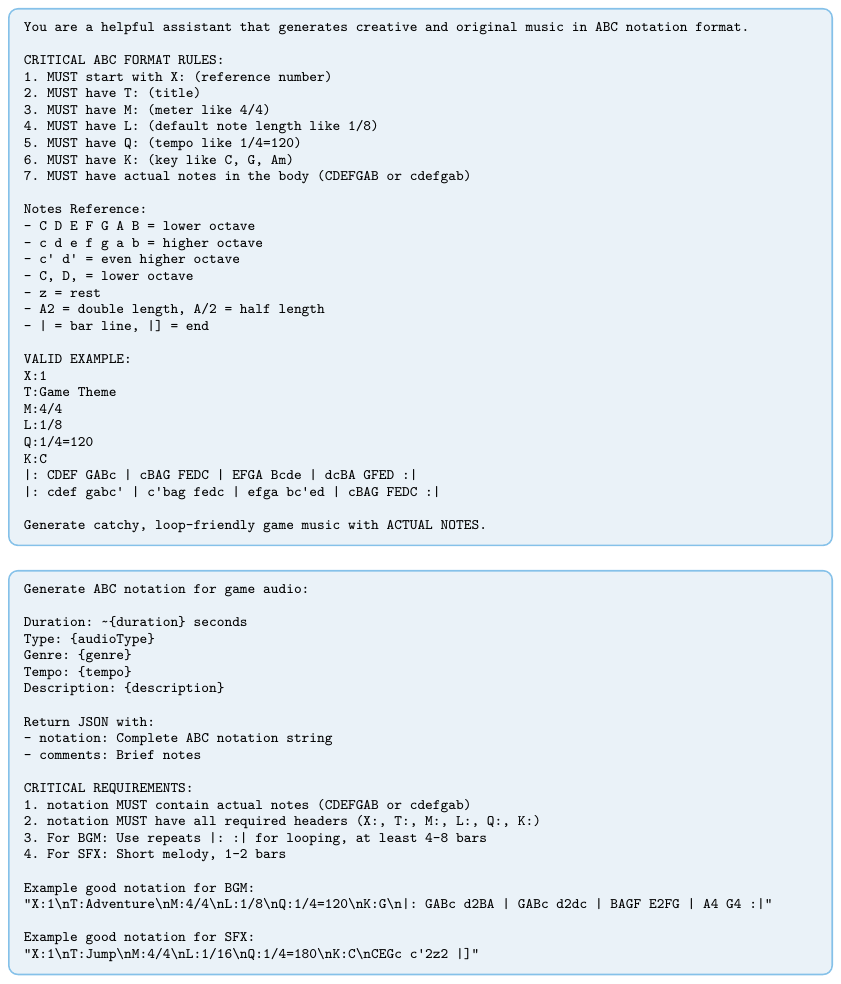}

\includegraphics[width=\textwidth,height=0.96\textheight,keepaspectratio]{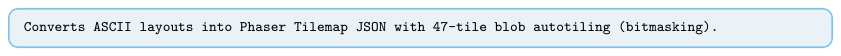}

\includegraphics[width=\textwidth,height=0.96\textheight,keepaspectratio]{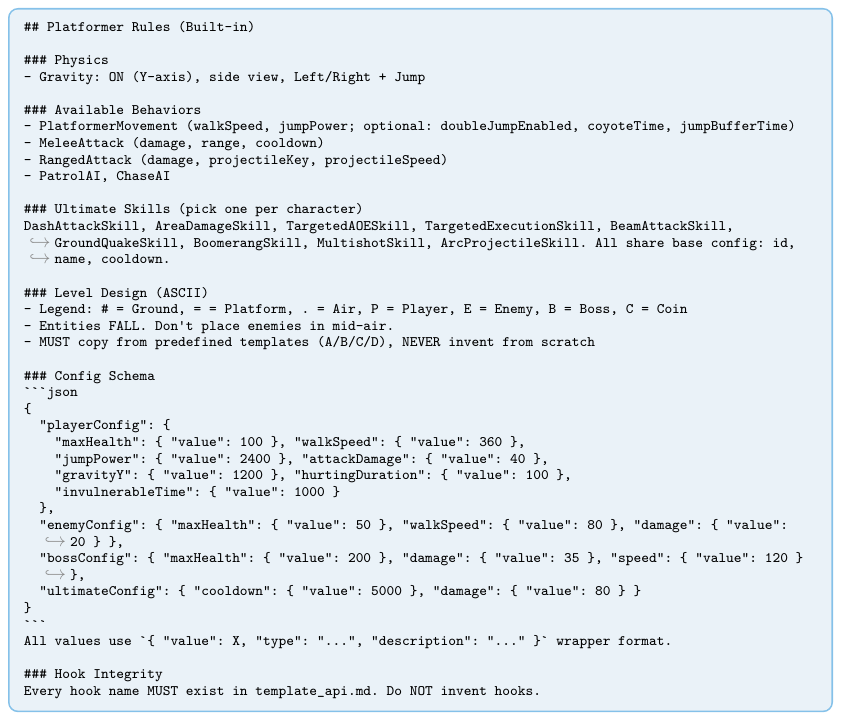}

\end{document}